# Space and Lunar Interferometry: Emerging Concepts and Pathways

Gioia Rau*[a]
a. Renaissance Philanthropy

## ABSTRACT

Space- and lunar-based interferometry are moving from long-range aspiration toward a concrete landscape of mission studies, pathfinders, and staged architecture concepts. This invited review surveys that landscape across two principal platform pathways: free-flying space interferometers and lunar-surface interferometers, with structurally connected designs, plus hybrid space-VLBI arrays that extend ground networks with a space element. Throughout, I trace how the science target sets the wavelength regime and angular resolution, which in turn fix the baseline, the architecture, and the dominant implementation risk.

The pathways span a variety of science: mid-infrared nulling concepts such as LIFE target temperate terrestrial exoplanets through thermal-emission spectroscopy; lunar far-side radio arrays (FARSIDE, FarView) open the low-frequency Universe from the only radio-quiet site in the inner Solar System; lunar UV/optical concepts such as the Artemis-enabled Stellar Imager (AeSI) and MoonLITE revisit very-long-baseline imaging of stellar surfaces and open a path to microarcsecond astrometry for the dynamical masses that direct-imaging target selection for the Habitable Worlds Observatory (HWO) requires; far-infrared interferometers (SPIRIT/SPICE), space VLBI (BHEX), and X-ray concepts extend the reach to planet formation, high-redshift galaxies, and black-hole physics; and lunar laser interferometry (LILA) applies the same platform to mid-band gravitational waves, measuring spacetime strain rather than angular position.

Across this wide range, the concepts share a set of enabling technologies: precision formation flying, absolute metrology, beam combination and nulling, cryogenics, and autonomous deployment – maturing largely as byproducts of flagship programs, a dynamic that increasingly governs which concepts become missions. I summarize the pathways in baseline, wavelength, science return, deployment strategy, enabling technology, and operational risk, and close with the decisions of the coming decade needed to turn these complementary concepts into operational observatories: pathfinder sequencing, the technology tall poles, and the funding models that can carry technologies from laboratory demonstration to flight readiness.

## 1. INTRODUCTION

The angular resolution of a single telescope is set by its aperture. NASA's planned Habitable Worlds Observatory (HWO), with a ~6-8 m mirror, will be transformative. Yet for any single-dish telescope this same aperture limit places certain science just out of reach, for example imaging the surfaces of other stars, or resolving the scales on which planets form. Interferometry lifts that limit by combining light from telescopes separated by tens to thousands of meters, synthesizing a virtual aperture far larger than any monolithic mirror, with angular resolution scaling roughly as wavelength divided by baseline. Kilometer-to-multi-kilometer baselines therefore open angular regimes no monolithic aperture can practically reach.

On the ground, interferometry is mature and productive: VLTI and CHARA routinely achieve milli-arcsecond imaging, including the first direct images of stellar surfaces beyond the Sun [65, 66], precision exoplanet orbits and atmospheres [67], and refined measurements of the cosmic distance scale [68], while advancing fringe tracking, beam combination, image reconstruction, and high-contrast methods [1-4]. But the atmosphere imposes a hard ceiling, making UV interferometry impossible from the ground and severely limiting optical/IR interferometry; Earth's atmosphere also limits sensitivity and calibration stability. Moving off-Earth removes that ceiling and opens wavelength windows inaccessible from the ground, while changing the trade space for thermal background, structural support, baseline stability, servicing, and operations: a space-based interferometer on a ~100 m baseline reaches milliarcsecond resolution at optical and UV wavelengths, sharp enough to resolve habitable-zone separations and sub-AU structure in the nearest systems.

*gioiarau.space@gmail.com

Mission studies for off-Earth interferometers are becoming concrete: quantitative yield and retrieval studies for free-flying mid-infrared nulling, anchored by the LIFE initiative [5-8]; NIAC and KISS reference studies for lunar UV/optical interferometry [9-11]; pathfinder payloads and deployable arrays for lunar far-side radio [12-15]; dedicated astronomy pathfinders – STARI, designed to perform the first in-space transfer of starlight between free-flying spacecraft, and SEIRIOS, aiming at the first stellar fringes in space [20, 34] – and in-flight demonstrations of formation flying and metrology [16-19], most recently ESA's PROBA-3 [18, 19], which after its December 2024 launch held a ~150 m formation to millimeter-level accuracy in 2025. The same platforms are now being considered for a fundamentally different measurement as well: lunar laser interferometry for mid-band gravitational waves [36].

This momentum also revisits long-standing unfinished business: NASA's Space Interferometry Mission (SIM) [58, 59], endorsed by earlier decadal surveys but never flown, was to deliver microarcsecond astrometry and the dynamical masses of nearby planetary systems – capabilities that remain only partially addressed and now feed directly into target preparation for HWO.

Different science questions map to different wavelengths, baselines, stability requirements, and infrastructure dependencies. Free-flying mid-infrared nulling is the strongest present pathway for temperate rocky exoplanets; lunar far-side radio is compelling because the far side is the only site in the inner Solar System that combines access to these frequencies otherwise blocked with shielding from terrestrial interference; and lunar UV/optical imaging is a newly credible, infrastructure-coupled pathway to sub-milliarcsecond science – resolving stellar surfaces and their magnetic activity, mass loss, AGN cores and accretion environments, and, at solar-system scales, the shapes and surfaces of asteroids and outer-system bodies – with the same platform identified by a recent KISS study as suited to microarcsecond astrometry for dynamical masses of direct-imaging targets [11]. The lunar surface supports a further, different-in-kind application as well: laser interferometry for mid-band gravitational waves, which measures spacetime strain rather than angular position.

Given the breadth and rapid evolution of the field, this review aims to be representative rather than comprehensive – illustrating each pathway with characteristic concepts rather than cataloguing every effort – and the omission of any particular study does not imply a judgment of its merit.

## 2. A SCIENCE-TO-ARCHITECTURE MAP

Off-Earth interferometry has often been framed around *where* the instrument flies – "a space interferometer," "a lunar interferometer". Organizing instead by what each concept is trying to measure makes the trade space legible: the science target sets the wavelength regime and the angular resolution required, those two together structure the baseline, and the baseline in turn drives the architecture (structurally-connected, free-flying, or surface-deployed) and with it the dominant implementation risks/challenges.

Structurally-connected concepts hold the apertures on a single rigid or deployable structure (a boom or truss) trading long, reconfigurable baselines for the elimination of formation flying; SIM [58], SPIRIT [21], and FKSI [44] are representative, and the approach remains the most flight-mature space-interferometry architecture. Free-flying concepts relax the baseline limit at the cost of precision formation control, while surface-deployed (lunar) concepts obtain long baselines from a stable platform at the cost of the surface environment.

Table 1 lays out the relating science drivers, wavelength regimes, baseline needs, candidate architectures, and the challenge that dominates each. The boundaries are deliberately soft, as several concepts serve more than one science case, a single science case is often addressed at more than one wavelength, and the most demanding technologies, for example absolute metrology, beam combination, formation control, and cryogenics, recur across Table 1 rows.

The field has diversified in its science while converging on a common set of enabling capabilities, which the rest of this review takes up pathway by pathway (Sections 3-4) and then as cross-cutting technology (Section 5).

**Table 1**. Representative off-Earth interferometry pathways and their primary science space.

| Pathway | Primary science | Wavelength regime | Representative baseline / scale | Dominant implementation challenge(s) |
|---|---|---|---|---|
| Free-flying MIR nulling | Nearby temperate exoplanets; thermal-emission spectra; biosignatures | 4-18.5 µm class | Tens to hundreds of m, adjustable | Null depth, formation control, cryogenic stability |
| Lunar UV/optical imaging | Stellar surfaces, magnetic activity, mass loss, AGN cores and accretion; solar-system small-body shapes | UV/visible/near -IR | 0.5-1 km first stages; expandable | Surface deployment, delay lines, contamination, dust |
| Lunar far-side radio | Dark Ages / Cosmic Dawn, solar bursts, exoplanet magnetospheres | Low-frequency radio (≈1-50 MHz) | Kilometer to hundreds-of-km arrays | Deployment, calibration, power, relay comms |
| Far-IR space interferometry | Protostellar/debris disks, high-redshift galaxies, cold dust | 25-400 µm class | Tens of m to kilometer-class | Cryogenic collectors, beam transport, detectors |
| Space VLBI / hybrid arrays | Photon rings, black-hole spin, jets, compact AGN | mm-cm radio | Earth-diameter-plus baselines | High-rate downlink, frequency stability, ground ops |
| X-ray interferometry | Black-hole coronae, AGN jets/tori, SMBH binaries | ~0.15-2.5 nm (0.5-8 keV) | ~1-100 m baseline; formation-flown ultra-long focal length (10^3-10^6 km) | grazing-incidence beam handling, formation metrology, photon efficiency, and technology immaturity |
| ***Lunar GW interferometry** | Mid-band gravitational waves: BH/NS merger early warning, $H_0$, neutron-star equation of state, GR tests | Mid-band GW, ≈0.1-10 Hz (strain, not EM) | ~tens of km arms (staged from a compact strainmeter) | Seismic isolation, suspended test masses, laser metrology, surface deployment |

*Measures spacetime strain not angular resolution; the wavelength and scale columns denote a GW frequency band and arm length (see Sect. 4.3).

Two axes organize the rest of this review. The first is platform – free-flying space interferometers (Section 3) and lunar-surface interferometers (Section 4), each surveyed concept by concept across wavelengths. The second is the enabling technology those concepts share (Section 5), which ultimately governs which of them become missions.

Figure 1 below places these pathways on common axes: for each, it shows the representative wavelength and baseline ranges – whose ratio sets the achievable angular resolution, θ ≈ λ/B – together with the deployment platform and an approximate architecture maturity – analogous to (though not a formal assessment of) NASA's technology-readiness level (TRL).

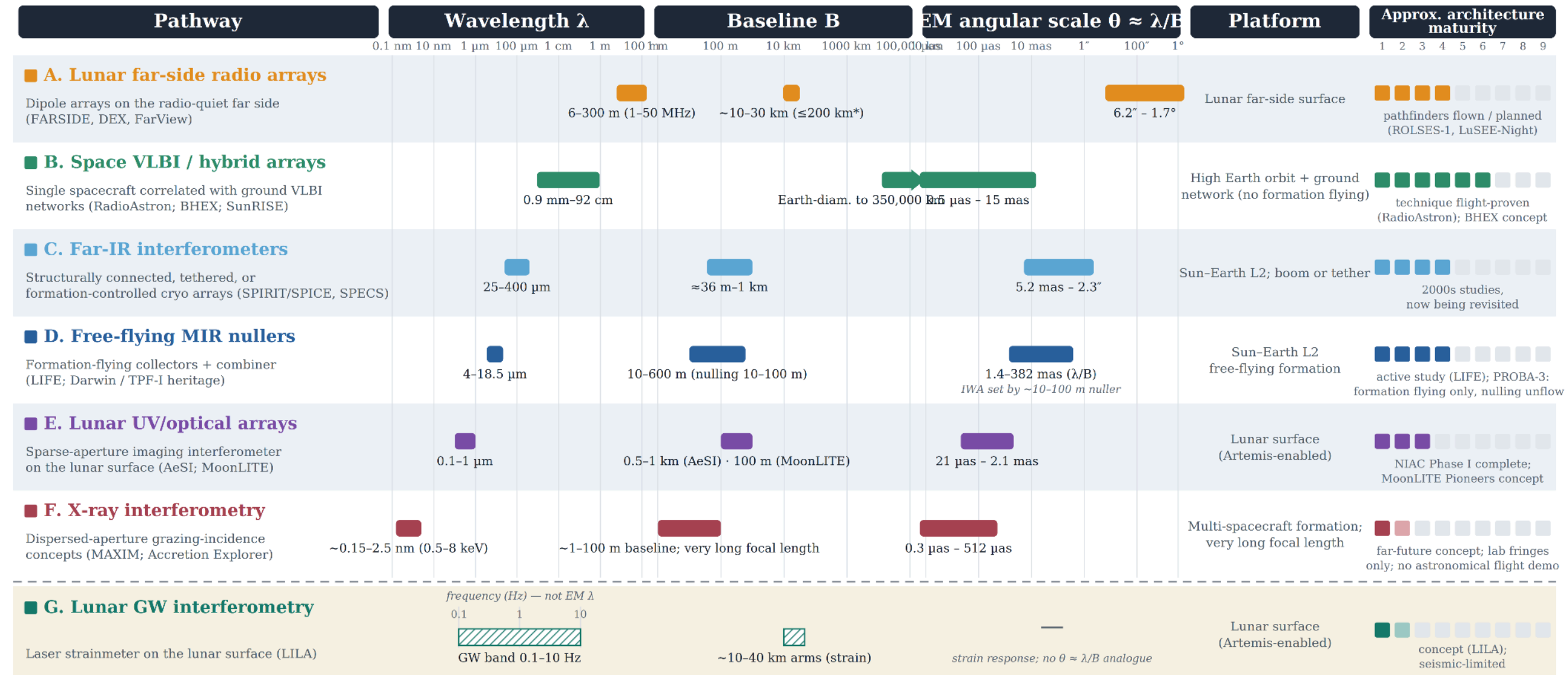

**Figure 1**. Representative off-Earth interferometry pathways across wavelength or frequency, baseline scale, angular-resolution scale, platform, and qualitative maturity.

**Notes**: Bars span representative ranges from the cited concept studies; wavelength and baseline axes are logarithmic. Filled boxes are qualitative architecture-readiness indicators, not formal NASA TRL assessments.

Row A baselines: ~10-30 km for FARSIDE / FarView-class reference designs; ~200 km appears in larger-site / ultra-long-wavelength trade studies. Row G is different in kind: LILA measures spacetime strain, not angular resolution - shown on its own frequency axis (Hz), with detector arm length in place of a resolving baseline, and no θ ≈ λ/B analogue (see Sect. 4.3).

Figure 2 complements Table 1 and Figure 1, placing the off-Earth interferometry concepts surveyed here in the two-dimensional space that most directly sets their architecture: observing wavelength against array baseline (or array scale). Concepts with otherwise unrelated science cluster by the baseline their target resolution demands, and the diagonal lines of constant angular resolution, θ ≈ λ/B, show how each reaches into the milli- and micro-arcsecond regimes. Ground-based optical/IR arrays are included for reference. The box extents are representative ranges drawn from the cited space concept studies and are intended for orientation rather than as formal specifications.

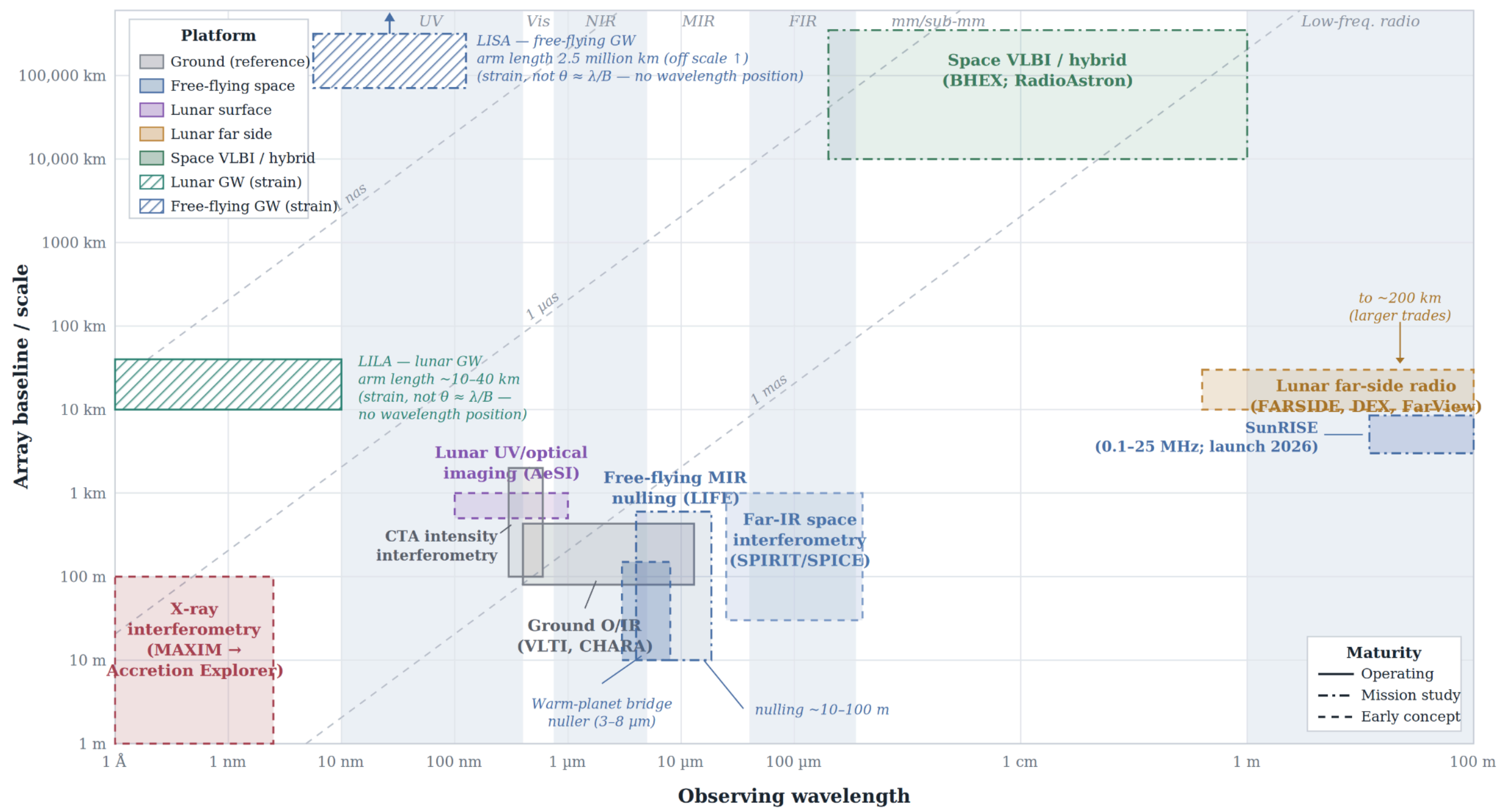

**Figure 2**. Schematic overview of off-Earth interferometry concepts in wavelength–baseline space. Each box spans the representative observing-wavelength and array-baseline (or array-scale) ranges of a pathway, color-coded by platform; border style indicates approximate maturity (operating, mission study, or early concept). Dashed diagonals are lines of constant angular resolution, $\theta \approx \lambda/B$. Box extent indicates representative wavelength and baseline ranges, not formal limits. Ground-based optical/IR arrays (VLTI, CHARA) are shown for reference. X-ray interferometers occupy short physical baselines (~1-100 m) despite their very long focal lengths; LILA measures strain, not angular resolution, and is shown for scale only (see Sect. 4.3).

## 3. FREE-FLYING SPACE INTERFEROMETRY

### 3.1 Mid-infrared nulling for exoplanets

In nulling interferometry, the apertures are combined so that on-axis starlight destructively interferes and is suppressed – the interferometric analogue of a coronagraph – revealing the faint thermal emission of off-axis planetary companions.

The free-flying Large Interferometer For Exoplanets (LIFE) initiative builds upon the heritage of the Darwin [42] and TPF-I [43] mission concepts, transitioning from early-stage conceptual advocacy to rigorous quantitative trade studies, with designs considering baselines of order ~10–600 m. Its yield study finds that a four-aperture, 2 m-class interferometer with 5% throughput over 4-18.5 µm could detect up to ~550 planets with SNR ≥ 7 in a 2.5-year search, including ~25-45 rocky planets in empirical habitable zones [5]. By increasing the collectors to 3.5 m, the yield rises to as many as ~770 planets, including roughly 60-80 rocky habitable-zone planets [5].

Thermal-emission spectra help constrain effective temperature, radius, and atmospheric species such as $CO_2$, $H_2O$, $O_3$, and $CH_4$, complementary to reflected-light missions such as HWO [6,7]. A dedicated LIFE candidate-target database, published in 2024, now supports these architecture trades and future target selection [29]. The architecture remains open: a five-telescope pentagonal kernel-nuller can outperform the classic four-telescope Emma X-array configuration (LIFE's baseline design) at fixed collecting area by about 23% for Earth-twin detections, provided that achromatic phase shifting can be implemented robustly [8].

Recent LIFE baseline-trade studies suggest that restricting the nulling-baseline range to 25-80 m, including to discrete or narrower baseline sets, may reduce operational complexity at only modest loss of performance (<10%), but note that tradeoffs between overall performance and implementation simplifications must be made [e.g., 32].

A nearer-term, structurally connected 'bridge' architecture, preceding a LIFE-class formation flyer, has long been advocated in the Fourier-Kelvin Stellar Interferometer (FKSI) lineage: a 3–8 µm, passively cooled nulling interferometer targeting warm planets (~600–1000 K, with FKSI studies centered near ~760 K equilibrium temperature), where the planet/star contrast is more favorable than at 10 µm [44, 45]. By observing the known population of radial-velocity planets, such a mission uses the interferometrically measured inclination to break the sin(i) degeneracy and recover true masses, together with radii and atmospheric constituents, leveraging two to three decades of radial-velocity monitoring of nearby M-to-F stars and reaching from hot Earths to warm sub-Saturns at roughly 0.05–2 au [44, 45]. The formation-flying PEGASE concept (1.5–6 µm) occupies a similar niche [46].

A complementary risk-reduction branch, explored in ESA-backed studies presented at SPIE in 2024-2025, examines a single-spacecraft nulling interferometer that sidesteps the hardest formation-flying problems; whether its reduced aperture and baseline still deliver a compelling science return remains open, but it signals that the community is no longer optimizing a single design lineage [31].

### 3.2 Far-infrared interferometry

Far-infrared interferometry trades angular-resolution extremes for sensitivity. SPIRIT [21] and its successor SPICE [47] were developed for imaging and spectroscopy at wavelengths background-limited from the ground, targeting protostellar and debris disks and high-redshift galaxies, delivering sub-arcsecond imaging and R ≈ 3000 spectroscopy from a two-telescope Michelson design.

For mid- and far-infrared interferometry, cooling the optics reduces telescope self-emission by roughly three orders of magnitude, moving the sensitivity limit off the warm-optics floor and down to the astrophysical background – reaching µJy to sub-µJy levels. This sensitivity gain, not merely the removal of atmospheric absorption, is one of the decisive motivations for space-based IR interferometry.

### 3.3 Space VLBI and hybrid architectures

Space VLBI offers another model for reaching micro-arcsecond resolution: a single spacecraft correlated with ground networks. As a heterodyne interferometer, BHEX would observe at high-frequency (submillimetre) radio wavelengths to reach photon-ring resolution around M87* and Sgr A* [22,23]. Rather than physically combining light beams in space, it records wave phases using an ultra-precise onboard atomic clock, leveraging the existing Event Horizon Telescope (EHT) ground network and an ultra-high-speed laser downlink to transmit the data for ground-based correlation rather than building a wholly new deep-space array.

More broadly, VLBI belongs to a class of interferometers that correlate independently recorded signals rather than optically combining light: a strategy that relaxes path-length and formation requirements at the cost of raw sensitivity. In the radio regime, this extends beyond BHEX to space arrays such as SunRISE, a six-CubeSat constellation forming a low-frequency (0.1–25 MHz) solar radio interferometer above the ionospheric cutoff [60], which demonstrates free-flying space interferometry at the frequencies that motivate the lunar far-side arrays of Sect. 4.2 – albeit for heliophysics and with relaxed positional requirements relative to optical work. Its optical counterpart, intensity interferometry (the Hanbury Brown–Twiss effect [64]), correlates second-order intensity fluctuations between independent apertures and is currently seeing a profound ground-based revival across major imaging atmospheric Cherenkov telescope arrays [61, 62] but has not yet been developed as an off-Earth concept. While historically sensitivity-limited to exceptionally bright targets (such as the landmark survey of 32 hot stars resolved down to ~0.4 mas by the Narrabri Stellar Intensity Interferometer [63]), modern photon-counting architectures tolerate arbitrarily long baselines with zero inter-telescope optical connections, making it a conceptually vital – though not yet off-Earth – member of this architectural family.

### 3.4 X-ray interferometry

X-ray free-flying interferometry, descending from MAXIM-class micro-arcsecond concepts [26], is being revisited through sparse-aperture, large free-flying X-ray architectures such as the Accretion Explorer, which targets milli- to micro-arcsecond imaging of black-hole coronae, relativistic jets, and supermassive-black-hole binaries at ~0.5-8 keV [38]. It marks the short-wavelength limiting case of angular resolution; however, it remains the least technically mature branch of space interferometry. To date, only laboratory fringe formation has been demonstrated, with no flown systems. This is owing to the strict demands of grazing-incidence optics, the exceptionally long focal lengths required, and the sub-nanometer formation metrology needed to hold meter-to-tens-of-meter baselines to the necessary precision.

## 4. LUNAR INTERFEROMETRY

Having surveyed space-based concepts – free-flying formations and their structurally connected and single-spacecraft variants – we turn to the lunar surface, the second principal platform. Free-flying and surface-deployed architectures are best understood as a trade in which each removes the other's hardest problem. Free-flying architectures offer arbitrarily long, reconfigurable baselines, a benign thermal environment at a quiet orbit, full-sky access, and independence from surface infrastructure, at the cost of precision formation flying. Structurally connected designs sidestep formation flying but cap baselines at boom or tether scale. Surface-deployed architectures eliminate formation flying entirely while retaining growable baselines, replacing active station-keeping with the passive geometric stability of a fixed surface, and add a buildable, serviceable platform and access to uniquely lunar sites (the radio-quiet far side and the seismically quiet, ultra-high-vacuum surface) – at the cost of dust, extreme thermal cycling, terrain, and dependence on the scaling of Artemis infrastructure. No single pathway dominates, which is why this review treats them as complementary rather than competing (a theme we return to in the roadmap of Sect. 7).

### 4.1 Lunar UV/optical interferometry

The Artemis-enabled Stellar Imager (AeSI) NIAC Phase I study [9, 10] represents the clearest recent formulation of a lunar UV/optical interferometer. It targets the resolved imaging of stellar surfaces and magnetically structured atmospheres – with profound implications for stellar dynamos and mass loss – while offering broader applicability to AGN cores, supernovae, planetary nebulae, and late stellar evolution. At Solar System scales, this resolution offers a "flyby without the flyby," resolving the shapes, surfaces, and binarity of hundreds of asteroids and outer-system bodies without dispatching dedicated spacecraft [11]. Utilizing a 1 km diameter array with an initial 15-element architecture – consisting of rover-deployed primary stations and central delay-line stations – AeSI is designed to expand to 30+ elements in staged deployments, serving as a stepping stone toward larger lunar and free-flying arrays across wavelengths [9, 10].

The lunar platform is attractive because vacuum operation removes the need for long enclosed delay-line tunnels, atmospheric coherence-time limits are eliminated, and the UV becomes accessible from off-Earth. Furthermore, a stable lunar surface relaxes the rigorous nanometer-level station-keeping and baseline-knowledge requirements inherent to free-flying configurations. However, the stability problem is transformed rather than eliminated, becoming an issue of dust-tolerant metrology, extreme thermal control, regolith contamination mitigation, autonomous surface deployment, and long-term maintainability. This architecture is also intrinsically dependent on the scaling of Artemis infrastructure for power and servicing. To bridge the gap between current technology and this grand vision, a nearer-term precursor has been proposed: MoonLITE [33]. As a CLPS-compatible concept, MoonLITE is designed to deliver the first separated-aperture optical interferometer beyond Earth using a single commercial landing. A rover would deploy a single 5 cm outboard telescope approximately 100 m from a matching inboard telescope on the lander, establishing a two-element, single-baseline observatory. The Moon's platform stability and long atmospheric-free coherence times make single-mode-fiber beam combination with long coherent integrations practical. MoonLITE is projected to reach roughly 1 mas resolution on targets as faint as 17th magnitude - acting as the single-baseline rung beneath AeSI's multi-element array. Even this precursor is highly productive; a 2025 KISS study notes that a two-element lunar interferometer could directly resolve brown-dwarf radii (~0.1 mas at 10 pc) that remain inaccessible from the ground [11].

A companion KISS study addressed the free-flying orbital case [57] – a reminder that these are complementary architectures, not a single choice to be made.

For imaging arrays, maximum baseline sets the nominal angular scale, but image fidelity is governed by the uv coverage (the range of baseline lengths and orientations the array samples), closure-phase precision (a calibration-robust phase combination formed around each triangle of apertures), calibration stability, and the number and geometry of collecting elements.

Beyond imaging and nulling, the lunar surface enables a third interferometric mode: high-precision astrometry. The same 2025 KISS study outlines a fixed-baseline lunar astrometric interferometer that could measure the masses and orbits of up to ~120 HWO candidate target stars at ~0.1 μas precision on a monthly cadence [11]. This specialized configuration would supply the definitive dynamical masses that direct-imaging target selection requires, filling a critical strategic gap that imaging and nulling alone cannot address. The same resolution reaches other compact targets: the nearest white dwarfs, Sirius B and Procyon B (~30 μas), would test white-dwarf mass-radius relations, and young pre-main-sequence stars in nearby star-forming regions would constrain stellar evolution at the earliest ages [11].

While a seminal 1996 study [41] originally favored free-flyers in the absence of lunar infrastructure, the advent of the Artemis era, the multi-track development of the ambitious AeSI roadmap, immediate commercial lander pathways via MoonLITE, and dedicated astrometric facilities argue that the timeline has shifted as lunar surface options have become much more competitive and timely. The lunar surface is now poised to be the optimal proving ground for next-generation UV/optical interferometry.

### 4.2 Lunar far-side radio interferometry

The radio-quiet lunar far side opens access to frequencies below ~30 MHz that are blocked by the terrestrial ionosphere and severely swamped by anthropogenic interference. Beyond the Dark Ages and Cosmic Dawn 21-cm signal, this window hosts exoplanetary and stellar magnetospheric radio emission (a probe of magnetic fields relevant to habitability), the low-frequency Sun and heliosphere, and cosmic-ray and plasma processes in the local Universe. To exploit this environment, the FARSIDE concept [12] proposes an array of 128 dual-polarization antennas distributed over a ~10 km footprint, deployed by four rovers in a four-arm spiral configuration and tethered to a central base station for processing, power, and data relay.

However, scaling these networks presents steep challenges; based on an ESA Concurrent Design Facility study, the Dark Ages Explorer (DEX) concludes that the immense antenna count required to isolate primary Dark Ages signals is not yet feasible with current high-TRL technology, though it outlines a viable medium-term development path [13]. To bypass the mass constraints of launching thousands of elements from Earth, FarView [14] extends this paradigm to a massive 100,000-dipole array, covering ~200 km², designed to be fabricated in-situ from lunar-regolith-derived materials over a 4-8 year campaign.

Near-term non-interferometric pathfinders are strategically vital to validating this roadmap. The LuSEE-Night mission [15] will target the pristine far-side spectrum from a static lander, relying on the Lunar Pathfinder satellite for data relay after the host lander's primary operations conclude. This mission serves as a critical operational reminder: the very isolation that grants the lunar far side its unparalleled radio-quiet advantage also transforms data communication into a complex orbital relay problem that must be resolved before large-scale interferometric arrays can be realized.

Beyond deployment and calibration, lunar far-side radio astronomy depends on protecting the radio-quiet zone itself. As Artemis-era communications, navigation, power systems, and commercial surface assets expand, spectrum-allocation policy, relay-satellite design, transmitter siting, and operational exclusion zones become part of the observatory architecture: without deliberate RFI governance, the far side's unique shielding from Earth could be compromised by local lunar infrastructure. Significant groundwork has been established by organizations such as UNOOSA [52]; sustained efforts are critical to ensuring that lunar sites of extraordinary scientific importance remain pristine [e.g., 53].

### 4.3 Gravitational-wave interferometry from the lunar surface

The lunar surface also enables a fundamentally different class of interferometer: one that measures spacetime strain rather than electromagnetic phase. The Laser Interferometer Lunar Antenna (LILA) [36] concept exploits the lunar environment to open the highly anticipated mid-band gravitational-wave window (~0.1-10 Hz). This "decihertz gap" lies squarely between ground-based detectors (such as LIGO, Virgo, and KAGRA, which are limited below ~10 Hz by terrestrial seismic and Newtonian noise) and the space-based LISA mission (optimized in the millihertz regime). Apollo-era and subsequent analyses indicate a much quieter seismic environment than Earth's at the relevant frequencies, though the detailed noise environment for a precision interferometer remains site-, frequency-, and deployment-dependent – motivating co-deployed seismometry to characterize site noise. This seismic quiet, combined with the Moon's ultra-high natural vacuum, provides the enabling environment for this architecture. Designed for a phased CLPS/Artemis deployment, LILA would progress from an initial lander-scale station using ultra-precise optomechanical readouts to monitor lunar resonant normal modes, toward a long-baseline "LILA-Horizon" configuration comprised of three surface stations forming a triangular array with ~40 km laser arms. Concurrently, the network would probe the Moon's deep interior structure with unprecedented geophysical sensitivity. LILA highlights the fact that the lunar surface's value as an interferometric platform extends well beyond traditional optical imaging.

The scientific yield of the decihertz band is distinct and highly complementary to other observatories. Sub-hertz sensitivity provides an "early warning" system by tracking stellar-mass black hole and neutron star binaries days to weeks ahead of their final merger, enabling real-time multi-messenger targeting that is otherwise impossible. Furthermore, decihertz measurements are uniquely positioned to sharpen constraints on the Hubble constant, probe the neutron-star equation of state, execute stringent tests of General Relativity, and map previously inaccessible sources, including white-dwarf mergers (Type Ia supernova progenitors) and intermediate-mass black hole tidal disruption events.

### 4.4 Lunar environmental and infrastructure constraints

The Moon is not simply a quiet version of free space. Surface systems must contend with brutal thermal cycling, abrasive dust, optomechanical contamination, limited sky visibility, challenging terrain, localized lander vibrations, power storage limitations, and complex communications geometry. For lunar surface arrays, site latitude, horizon, solar illumination, thermal constraints, power, Earth visibility/relay, and target dwell time matter.
Current NASA environmental models span roughly +250 °F to −410 °F [48], and the electrostatically charged regolith dust is highly abrasive, clingy, and notoriously prone to degrading thermal-control surfaces and optical coatings [37]; though mitigation approaches (protective covers, electrodynamic dust shields, operational protocols) are actively maturing.
However, two empirical benchmarks temper the most pessimistic expectations. First, the Chang'e-3 lander operated a small-aperture robotic Lunar Ultraviolet Telescope (LUT) on the surface, demonstrating stable photometric performance over roughly 18 months with no evidence of dust-induced degradation [27]. This was achieved by keeping the instrument enclosure closed at sunrise and sunset, when terminator dust is actively lofted by electrostatic charging.

Second, the Apollo 12 descent stage and the Surveyor 3 lander have maintained a separation of roughly 180 meters for over fifty years without any active intervention [11]. This demonstrates that the lunar surface provides exceptional passive geometric persistence "for free" - removing the active station-keeping burden inherent to free-flying arrays. It is worth noting, however, that this stability speaks only to the position of the macro-structures, leaving the nanometer-scale optical-path metrology and thermal control, that set actual performance, to be resolved by the payload design.
The ambient seismic environment is similarly benign. Recent findings confirm the Moon's ground motion--originally assumed to be a major obstacle in early 1996-era assessments of lunar astronomy--to be a minor issue for optical interferometry [11]. It is precisely this seismic quiet that makes the Moon highly attractive for the surface gravitational-wave interferometry discussed in Sect. 4.3.

Ultimately, the platform remains highly compelling because the lack of an atmosphere eliminates the need for long, enclosed vacuum delay lines and radically extends coherence parameters. Where a ground-based interferometer is strictly bounded by an atmospheric coherence volume of order $10^4$ $m^3$ and an isoplanatic angle of arcseconds, in the absence of atmospheric turbulence, the lunar surface removes the atmospheric coherence volume and isoplanatic-angle limits that constrain interferometry from Earth.

Where ground interferometers are limited to millisecond-scale coherent integrations by atmospheric turbulence, removing the atmosphere lifts that limit, enabling far longer coherent integration and thus access to substantially fainter targets, bounded only by instrumental metrology and thermal stability.

The stability problem is not eliminated, but transformed: performance limits are dictated by the instrument itself and its operations: its internal metrology, thermal management, contamination mitigation, and autonomous deployment--rather than by the sky.

## 5. ENABLING TECHNOLOGIES

The concepts surveyed above diverge widely in science and architecture, yet they converge on a set of enabling capabilities. Mapping architectures to capabilities to enabling technologies unites these concepts under a common ontology, and motivates cooperative work: the same technologies recur across mission concepts and wavelengths, from a free-flying mid-infrared nuller to a lunar UV/optical array to a far-infrared imager.

Figure 3 makes this mapping explicit, and the subsections below take the shared technologies in turn: formation flying, metrology, and control (Sect. 5.1); beam combination, nulling, and photonics (Sect. 5.2); cryogenic optics, processing, and precursor science (Sect. 5.3) – and close with the maturation priorities that gate the field (Sect. 5.4).

| | SCIENCE GOALS | | | | | | | | ENABLING TECHNOLOGIES | | | | | |
|---|---|---|---|---|---|---|---|---|---|---|---|---|---|---|
| | Exoplanet biosignatures (thermal-IR) | Stellar surfaces & activity | Planet formation / disks | AGN & black-hole cores | Photon ring / strong-field GR | Dark Ages 21-cm | Mid-band gravitational waves | High-z galaxies / cold dust | Formation flying | Absolute metrology | Nulling / beam combination | Cryogenic optics | Timing / correlation / data arch. | Deployment / autonomy |
| **LIFE** *Free-flying MIR nulling* | ● | | ○ | | | | | | ● | ● | ● | ● | | ○ |
| **SPIRIT / SPICE / SPECS** *Far-IR interferometry* | | | ● | | | | | ● | | ○ | ● | ● | | ○ |
| **BHEX** *Space VLBI* | | | | ● | ● | | | | | ● | | ○ | ● | |
| **MAXIM / Accretion Explorer** *X-ray interferometry* | | ○ | | ● | ● | | | | ● | ● | | | | ○ |
| **AeSI** *Lunar UV/optical* | | ● | ○ | ○ | | | | | | ● | ● | | ○ | ● |
| **MoonLITE** *Lunar UV/optical · precursor* | | ● | | ○ | | | | | | ○ | ● | | | ● |
| **FARSIDE / FarView** *Lunar far-side radio* | | | | | | ● | | | | | | | ● | ● |
| **LILA** *Lunar gravitational-wave* | | | | | | | ● | | | ● | | | | ● |
| **Pathfinders** *STARI · SILVIA · SEIRIOS · SunRISE* | | ○ | | | | | | | ● | ● | ○ | | ● | ○ |

● primary capability ○ secondary / partial *marker color = mission pathway*

**Figure 3.** Capability matrix: space mission concepts mapped to complementary science goals and shared enabling technologies. Filled circles are primary capabilities, while open circles represent secondary capabilities. Mission pathways are color-coded.

### 5.1 Formation flying, metrology, and control

Despite compelling science cases, no off-Earth interferometer has yet flown. The barrier has been system-level complexity and integration risk, not any single missing component, compounded by the absence of a dedicated flight line.

Precision formation flying and metrology are shared requirements. GRACE-FO's [49] Laser Ranging Interferometer measured spacecraft separations of ~220 km with ~nm/√Hz noise [17]; LISA Pathfinder demonstrated picometer interferometric metrology and drag-free control [16]; and PROBA-3 held its ~150 m formation to ~1 mm in both range and lateral position, autonomously [18,19].

These demonstrations address inter-spacecraft ranging and formation control; the final stage (matching optical paths to a fraction of a wavelength) additionally requires short, space-qualified optical delay lines, needed by free-flying combiners and lunar surface arrays (e.g., AeSI's final phasing) alike.

A complementary thread is the miniaturization of LISA Pathfinder-heritage inertial sensing: gravitational reference sensors – free-floating test masses that measure all non-gravitational accelerations of the host spacecraft – are being adapted to small-satellite scales [69, 70, 71], which would give each element of a formation its own inertial reference, enabling drift to be sensed and compensated in feed-forward rather than measured only relatively. Combined with inter-spacecraft laser metrology, such sensors could substantially relax the control burden of holding optical paths at the nanometre level.

STARI targets a 2029 launch: two propulsive CubeSats fly in formation at separations up to ~100 m in low Earth orbit, one relaying a stellar beam to the other, where an off-axis parabola reimages it into a single-mode fiber, which spatially filters the beam to a clean wavefront for high-contrast combination. Formation control uses differential GPS, LED beacons, and fast-steering mirrors in closed loop, explicitly to retire optical and control risks for missions such as LIFE. SILVIA [24], a three-spacecraft, ~100 m demonstration of sub-micrometre relative control in low Earth orbit, is in JAXA Pre-Phase A2, and a companion JAXA concept, SEIRIOS [34] (a ~40 kg nanosatellite flying with two 6U CubeSats at 10-100 m baselines), aims to record the first stellar interference fringes ever obtained in space. These concepts build on distributed-aperture precursors such as VISORS [35], whose twin-CubeSat formation control at ~40 m separation provides immediate GNC heritage for STARI's starlight-transfer demonstration.

Crucially, these capabilities are maturing largely as technological byproducts of other flagship programs, such as JWST's deployable cryogenic optics, HWO's picometer-scale wavefront control, and advanced coronagraph metrology--meaning that interferometry's effective technology readiness level (TRL) is rising even in the absence of a dedicated astronomical flight. The ultimate cross-cutting tall pole remains absolute metrology, where tolerance thresholds scale strictly with wavelength and science mode. The longer mid-infrared wavelengths make nulling configurations comparatively forgiving in terms of absolute path knowledge (though deep nulls still demand tight differential-OPD stability), whereas resolved visible/UV imaging pushes toward a strict ~10 nm absolute metrology threshold that has not yet been demonstrated in flight.

### 5.2 Beam combination, nulling, and photonics

Beam combination sets sensitivity, observables, and – for nullers – the achievable starlight suppression. The Asgard/NOTT L′ -band nuller at the VLTI is building the high-contrast workflow space missions will inherit, targeting young planets and exozodiacal dust with photonic beam combiners and self-calibration [4]. Its recent four-telescope integrated-optics combiner reported the first broadband L′ deep-null measurements – an average raw null near $8\times10^{-3}$ and a self-calibrated null near $1\times10^{-3}$ at room temperature – with cryogenic testing identified as the immediate next step [28].

### 5.3 Cryogenic systems, onboard processing, and data architecture

Mid- and far-infrared interferometry depend fundamentally on stringent cryogenic stability. The SPIRIT mission concept [21] remains highly instructive because it treated cryogenics as an architectural baseline, pairing dedicated mechanical cryocoolers with a continuous adiabatic-demagnetization refrigerator (CADR) rejected at a 4 K structural interface to cool detectors to sub-Kelvin operating thresholds.

Furthermore, as space-based arrays scale up in element count, real-time onboard fringe tracking, autonomous phase calibration, and high-cadence data processing transition from operational efficiencies to baseline enabling technologies.

This is especially true for space VLBI and large distributed arrays, where the data system is not a downstream implementation detail but an architecture driver. Space VLBI requires stable time and frequency references, phase transfer, high-rate recording, and downlink capacity sufficient for ground correlation; BHEX-like concepts therefore depend as much on clock stability and laser communications as on collecting area. Lunar far-side radio arrays face a related but distinct scaling problem: large-N dipole fields will require onboard or local correlation, compression, calibration, RFI identification and removal [56], and data triage before relay, because raw-voltage downlink from thousands to hundreds of thousands of elements is not realistic. AI-enabled autonomy is likely to become part of this architecture: machine-learning methods are already being developed for radio-interferometric calibration workflows, automated VLBI calibration, and RFI identification, and analogous onboard systems could prioritize scientifically valuable data, flag corrupted streams, monitor instrumental state, and reduce telemetry volume before transmission [54, 55]. These methods should be treated as decision-support and compression layers; the underlying interferometric requirements remain: telemetry, time standards, onboard processing, correlation strategy, and calibration architecture must still be designed with the observatory from the start rather than added after the optical or antenna layout is fixed.

Simultaneously, precursor science must be treated as critical risk retirement rather than a mere preliminary exercise: NASA's ExEP Science Gap List explicitly flags habitable-zone exozodiacal dust as a high-priority strategic gap [30]. Recent modeling warns that hot and warm exozodiacal dust structures can severely degrade the contrast sensitivity, inflate integration times, and complicate noise budgeting for future exo-Earth direct-imaging missions such as the HWO [39,40]. This reality elevates near-term mid-infrared interferometric exozodi surveys from independent science investigations to mission-critical screening tools necessary for establishing the final target lists of upcoming flagship observatories.

### 5.4 Maturation priorities

If these mission concepts are to successfully transition into flight missions, a definitive list of technological "long poles" must be retired, and they are not equally hard. The primary technical hurdle is absolute metrology in the visible and ultraviolet regimes. Mid-infrared operation relaxes the wavelength-scaled optical-path requirements relative to visible/UV interferometry: micron-level path control may be sufficient for coarse fringe acquisition, but deep nulling still requires differential optical-path stability at the nanometer-to-tens-of-nanometers level, together with tight control of amplitude, polarization, and chromatic phase errors. Visible/UV imaging is less forgiving still, pushing toward ~10 nm-class absolute optical-path metrology for resolved stellar-surface imaging; this level of end-to-end performance has not yet been demonstrated in flight and remains a direct gate for AeSI/Stellar Imager-class architectures.

Second is system-level formation flying paired with end-to-end astronomical fringe acquisition. The constituent subsystems have each been validated separately: PROBA-3 at the millimeter level [18, 19], GRACE-FO at the nanometer level [17], and LISA Pathfinder at the picometer level [16]; but coherently combining starlight across independent, free-flying spacecraft remains unproven. This is precisely the critical operational gap that STARI and SEIRIOS are designed to close ahead of larger flagships like LIFE and any future space-based Planet Formation Imager (PFI) [3].

Third is deep, stable, cryogenic nulling. The NOTT ground instrument has demonstrated self-calibrated broadband nulls near $10^{-3}$ on sky at room temperature [28], while the LIFE consortium's Nulling Interferometry Cryogenic Experiment (NICE) at ETH Zürich has reached the required $<10^{-5}$ null depth on its ambient precursor bench at a single mid-infrared wavelength, with the transition to cryogenic (~15 K) operation now under way [72]. The remaining gap is therefore integration: combining that depth with broadband coverage, high throughput, and cryogenic stability in integrated-optics beam combiners equipped with robust achromatic phase shifting.

Fourth is the cryogenic chain required for the far-infrared, necessitating cold collectors, low-temperature detectors, and stable delay lines of the SPIRIT class [21].

Fifth, and specific to the lunar pathways, is surface deployment and autonomy. This includes rover-deployed baselines (MoonLITE, AeSI), in-situ deployment, and subsequent scaling, of massive arrays (FARSIDE, FarView), dust and thermal mitigation, and far-side orbital communications relays - with the added complexities of passive seismic isolation and the proposed 'GravComb' quantum-enhanced laser frequency comb metrology as the added requirements for LILA [36].

For the radio and VLBI branches, data architecture is itself a maturation priority: stable frequency standards, phase transfer, onboard correlation or compression, and high-rate relay/downlink must mature alongside the collecting hardware. NASA's SunRISE constellation is an early in-flight demonstration of exactly this chain – six free-flying CubeSats forming a low-frequency space interferometer, enabled by software-defined radios and precision GPS timing and positioning [60] – and its heritage bears directly on the correlation and relay architectures the lunar far-side arrays will require.

Underlying all of these domains, real-time onboard fringe tracking, system calibration, and autonomous operations must be designed to scale efficiently with element count rather than remaining bespoke implementations. The encouraging pattern is that the majority of these capabilities are actively maturing as technological byproducts of the JWST, HWO, and advanced coronagraph programs.

The decisive near-term move for the community is to integrate these demonstrated subsystems into targeted, end-to-end astronomical demonstrations, validating the free-flying and lunar surface tracks in parallel, before committing to a flagship flight line.

## 6. RELATIONSHIP TO GROUND INTERFEROMETRY AND LARGE-APERTURE FACILITIES

Space-based interferometers complement rather than replace ground facilities, which occupy two ends of a trade. The CHARA Array holds the longest optical baselines, up to 331 m, achieving angular resolutions of 0.20 milliarcseconds (mas) in the visible and 0.7 mas in the K-band with 1-m-class telescopes, resolving, for example, the gravity-darkened, rotationally distorted surface of the rapid rotator Altair [65]. ESO's VLTI leverages collecting area and sensitivity: its four 8-m Unit Telescopes, combined with dual-field phase referencing (now extended by GRAVITY-Wide to faint targets over wider fields [73]) reach objects far beyond CHARA's magnitude limits, while its movable 1.8-m Auxiliary Telescopes provide reconfigurable baselines up to 200 m, with resolutions from ~2 mas in the near-infrared to ~12.5 mas in the mid-infrared. Together they bracket the resolution–sensitivity trade that any off-Earth successor must improve upon.

These facilities make the full interferometric tool-chain operationally real: fringe tracking, beam combination, calibration, image reconstruction, and high-contrast observing; even though atmospheric turbulence and thermal background ultimately limit what can be done from the ground. They also remain essential development environments: current and planned upgrades broaden sensitivity, sky coverage, and observing modes, while instruments such as MATISSE, GRAVITY/GRAVITY+, NOTT, and LBTI/HOSTS [40] provide the calibration, nulling, and exozodiacal-dust measurements that future space architectures will inherit. The connection to flagship preparation is already direct: an ongoing NASA-funded program is using CHARA and VLTI observations to characterize HWO candidate target stars in support of direct-imaging target selection (R. Roettenbacher 2026, priv. comm.). The 2024 PFI update [3] still describes a >1.2 km ground array as transformational for planet-formation science, while also noting that emerging space-interferometry architectures could make a space-based PFI increasingly compelling and could serve as a pathway toward LIFE.

The relationship to the planned HWO and the coming generation of extremely large telescopes is similarly complementary. ESO's 39 m Extremely Large Telescope, now under construction, and the planned 25–30 m Giant Magellan Telescope and Thirty Meter Telescope provide immense collecting area and tens of milliarcsecond resolution for exoplanets, stellar populations, black holes, and time-domain astrophysics. HWO, designed as a 6-8-m-class space telescope flagship, targets the reflected-light characterization of nearby habitable-zone planets and broader ultraviolet, optical, and near-infrared astrophysics.
HWO and LIFE-like missions are therefore strongest as complementary observatories. Reflected-light and thermal-emission measurements probe different components of the planetary state vector, and joint retrievals dramatically reduce deep degeneracies among radius, albedo, temperature, composition, and climate state [7].

## 7. ROADMAP AND KEY DECISIONS FOR THE COMING DECADE

The concepts surveyed in this review are distinct in science and architecture, yet they rest on a shared interferometric logic, and, as Section 5 showed, on a shared technology base. The decade ahead will therefore be defined less by selecting a single 'winner' than by strategic choices that require international and cross-sector collaboration. Six stand out:

1) First, fly at least one end-to-end astronomical formation-flying pathfinder. The constituent subsystems have largely been demonstrated in isolation: ESA's PROBA-3 has retired much of the precision station-keeping and relative metrology risks, and STARI is designed to retire starlight acquisition, beam steering, and single-mode-fiber injection. However, no funded mission has yet combined independent beams of starlight into actual optical fringes in space. That step, achieving "first stellar fringes from separated free-flying spacecraft in space," is the single clearest gate on the path to any free-flying larger mission. The primary risk to the field is the continued deferral of integrating these demonstrated subsystems into a real mission [25].
2) Second, even though many propose to resolve "the free-flyer versus lunar-surface fork", I'd like to propose here a different approach: start small on each end, pursuing small-scale risk reduction on both frontiers: demonstrating initial feasibility and first fringes, through a rapid-iteration paradigm before scaling: fail fast, and build on what works, assuming a startup mentality. Frequency is the key: increase flight frequency via low-cost precursors, to accelerate pathfinder demonstrators; this accelerates programmatic maturity more effectively than deferring progress in anticipation of a multi-billion-dollar flagship mission. A single-baseline CLPS-enabled lunar precursor such as MoonLITE [33] is a natural early test example: a funded, commercially-delivered demonstration would validate the lunar surface approach for the larger arrays; whereas its programmatic absence leaves the free-flying path as the default baseline. A further argument for this incremental approach is epistemic. Cost growth through redesign is endemic to large space missions of every kind – but monolithic telescopes at least inherit calibrated cost models from flown predecessors. No astronomical space interferometer has ever flown, so estimates for formation-flying observatories rest on models with no reference class at all. Even SIM – its technology development complete and its costs independently assessed for the 2010 decadal survey [58, 59] – did not secure a ranking, and its cancellation left the field without a flown cost anchor. Small, flown precursors therefore do double duty: they retire engineering risk, and they establish the first real cost and schedule anchors on which credible flagship estimates can be built.
3) Third, keep lunar radio and lunar optical on separate maturity tracks. The two diverge because their pacing dependency differs: lunar far-side radio rests on a quiet-environment science case with pathfinders, array concepts, and relay logic already largely in hand; whereas lunar optical interferometry depends more heavily on the long-term buildout of sustained lunar surface infrastructure, including power, mobility, and human/robotic servicing for deployment and operation. Applying the same incremental approach outlined in #2 above, could rapidly accelerate maturity level for the far-side radio interferometry missions, unlocking unique science such as low-frequency cosmology that is physically unfeasible anywhere else in the inner Solar System.

4) Fourth, treat the two cross-cutting technology tall poles as the real near-term investment decision. (1) Absolute metrology at the ten-nanometer to tens-of-picometer level gates the visible/UV imaging and astrometric concepts; while (2) cryogenic deep nulling (targeting raw null depths of order $10^{-5}$ and effective depths approaching $10^{-8}$ via phase chopping) gates the mid-infrared nuller and therefore the characterization of temperate worlds. Because these diverse concepts share fundamental formation-flying and metrology gaps, a coordinated technology-maturation program will retire risk across the whole range far more efficiently than funding each mission concept's development in isolation.
5) Fifth, recognize that near-term progress is often paced less by science or technology readiness than by opportunity availability. Several of the concepts reviewed here are mature enough to be proposed, but lack a competed flight line at the appropriate scale and cadence; the lunar precursors in particular have at times had no open, suitably-scoped federal funding path. Aligning concept readiness with the actual programmatic sequence of Probe, Explorer, and small-mission opportunities is itself a vector the community must actively influence through structured advocacy and through keeping concepts and payloads proposal-ready. NASA's ASTRA Initiative [50] represents an example of a step forward in bridging this implementation gap.
6) Sixth, treat international coordination as an explicit strategic choice. Space interferometry has transitioned into a genuinely multi-agency endeavor: JAXA is advancing dedicated formation-flying demonstrations [24, 34]; ESA maintains key mid-infrared exoplanet and X-ray initiatives alongside its precision formation-flying flight heritage; the United States leads on lunar-surface arrays and small pathfinders; the United Kingdom has formally entered the landscape via its recent feasibility white papers (e.g., SPIFF [51]); and China is advancing substantial interferometric investments, including the TianQin and Taiji space-based gravitational-wave programs, the Discovering the Sky at the Longest Wavelengths (DSL) radio array, or its Lunar Orbital VLBI Experiment (LOVEX) aboard the Queqiao-2 relay satellite as part of the Chang'e-7 architecture. The decision facing the community is whether to consolidate these global assets into a shared metrology framework with modular component contributions, or to run duplicative national demonstrations that collectively slow the field.

Underlying all six choices, these off-Earth facilities extend the reach of ground-based facilities, and future space-based ones, each populating a unique region of wavelength, baseline, and angular-resolution space, complementary to the others. Seen as a portfolio, the decisions of the coming decade come down to sequencing a coherent set of complementary capabilities – integrating the demonstrated pieces, on the ground and off the Earth, before the window to do so closes; a schematic staging of these six choices across the decade is sketched in Appendix A (Fig. A1).

### 7.1 Funding Models

Related to opportunity availability is the question of who funds the steps between lab demonstration and full flight program. Much of space interferometry occupies a programmatic valley: too mature for research grants, yet too early for flight programs, cross-cutting enough that it belongs cleanly to no single mission line, and carrying a number of risks.

Government agencies have historically remained the ultimate route for flagship missions, and the decisions above are ultimately theirs to enable; but the integration steps that retire system-level risk, for example an end-to-end fringe demonstration, a single-baseline lunar precursor, a shared metrology or nulling testbed, are exactly the kind of bounded, higher-risk investments that philanthropic and private capital can execute on significantly shorter timescales than competed federal lines allow.

Recent astronomy offers clear precedent: privately and philanthropically supported efforts have stood up major facilities and convened the community studies and frameworks that seeded subsequent programs. Notable examples include the early philanthropic seed capital that financed the mirror casting for the Vera C. Rubin Observatory before its selection as a national decadal priority; the foundational wide-field data frameworks established by the Sloan Digital Sky Survey; and the recent Breakthrough Watch NEAR campaign on the VLT, which privately funded advanced mid-infrared coronagraphy to retire high-contrast exoplanet imaging risks. Used this way – as catalytic, early de-risking, and complementary to agency investment rather than a substitute for federal funding – such approaches and funding mechanisms could radically accelerate the technical maturation to make space interferometry concepts proposal-ready when the opportunities open.

## 8. CONCLUSIONS

Space and lunar interferometry offer a unique and unparalleled range of opportunities for the next generation of high-angular-resolution observatories. As surveyed in this review, the different architectural tracks populate complementary regions of discovery space: free-flying mid-infrared nulling holds the strongest near-term case for the atmospheric characterization of temperate exoplanets; lunar far-side radio arrays possess the clearest platform-specific advantage and a structured pathfinder ladder for unique low-frequency cosmology; lunar UV/optical interferometry enjoys renewed strategic relevance through its direct coupling to emerging surface infrastructure and operations; and far-infrared direct-detection and space-VLBI heterodyne concepts broaden the astrophysical scope to include planet formation, high-redshift galaxy evolution, and high-fidelity black-hole physics. The overarching task for the coming decade is to methodically sequence and integrate the core technologies: precision formation flying deep cryogenic nulling, nanometer-to-picometer-class metrology, photonic beam combination, autonomous lunar deployment, and the timing, correlation, and data architectures of the radio and VLBI branches; this will turn these complementary mission concepts into operational space observatories.

The programmatic landscape of the late 2020s has shifted the feasibility calculus that stalled earlier space interferometry initiatives. The historical barriers of extreme launch mass constraints, prohibitive mission costs, and restricted deep-space operations are actively being dismantled by the rise of commercial heavy-lift launch vehicles, the logistical regularity of Commercial Lunar Payload Services (CLPS), and the emerging infrastructure of the Artemis era, with philanthropic and private capital increasingly positioned to fund the early, high-risk steps. Rather than waiting for a single, multi-billion-dollar flagship line to carry the field forward, the modern era enables an agile, multi-track prototyping strategy where small-scale, high-cadence pathfinders can retire engineering risks in parallel. The precedent is instructive: NASA's Great Observatories demonstrated that complementary observatories, sequenced as a portfolio, can exceed the sum of their parts – a vision revisited for the interferometric pathways of this review in Appendix A (Fig. A2).

Ultimately, space and lunar interferometry represent a natural next frontier of observational astrophysics. As single-aperture space telescopes approach their practical engineering and physical scaling limits, humanity's pursuit of high-contrast and high-angular resolution astronomy must transition from an aperture-limited paradigm to a baseline-driven architecture. Embracing this era of multi-aperture systems allows us to scale up our observational baseline, overcoming past constraints on angular resolution and cosmic discovery space. By leveraging the virtually unbounded coherence parameters of the lunar vacuum and the vast baselines of distributed space formations, off-Earth interferometers will bridge the critical observational gaps left by single apertures: moving us beyond simple point-source detection and into an era where we can directly image the active surfaces of distant stars, map the underlying physics of the first cosmic structures, listen to the gravitational-wave sky in a band no other observatory can reach, and contribute decisively to assessing the habitability of nearby worlds.

## ACKNOWLEDGMENTS

The author thanks John Monnier, Kenneth Carpenter, and Jon Hulberg for helpful comments.

# APPENDIX A

This appendix collects two schematic figures that summarize, in deliberately informal visual form, the strategic argument of Sections 7-8.

Figure A1 stages the six choices of Sect. 7 across the coming decade; Figure A2 places the surveyed pathways in the visual tradition of the hand-drawn illustration that helped introduce NASA's Great Observatories program – a reminder that complementary observatories, sequenced as a portfolio, have precedent.

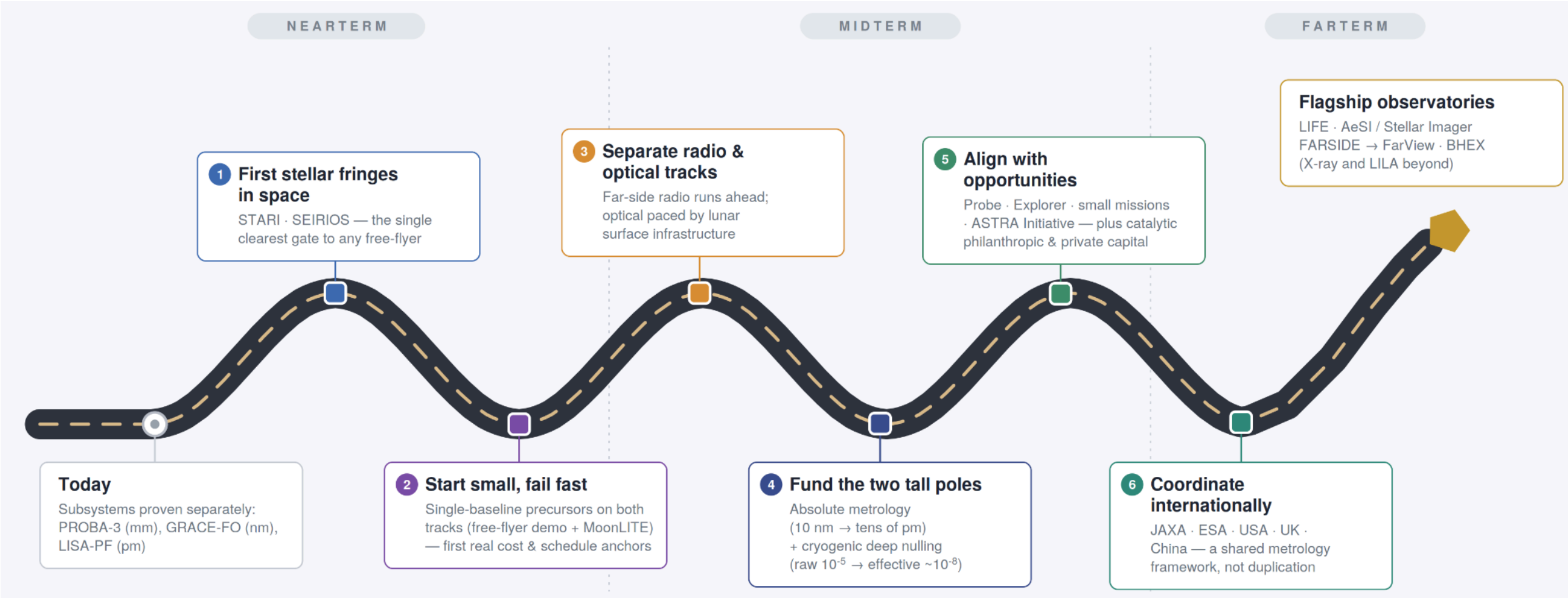

**Figure A1**. Schematic staging of the six strategic choices of Sect. 7 across near-, mid-, and far-term horizons, from subsystem demonstrations already flown (PROBA-3, GRACE-FO, LISA Pathfinder) through pathfinders and technology tall poles to flagship-class observatories. The ordering is indicative of dependency, not a programmatic timeline; milestones and thresholds are drawn from the studies cited in the text.

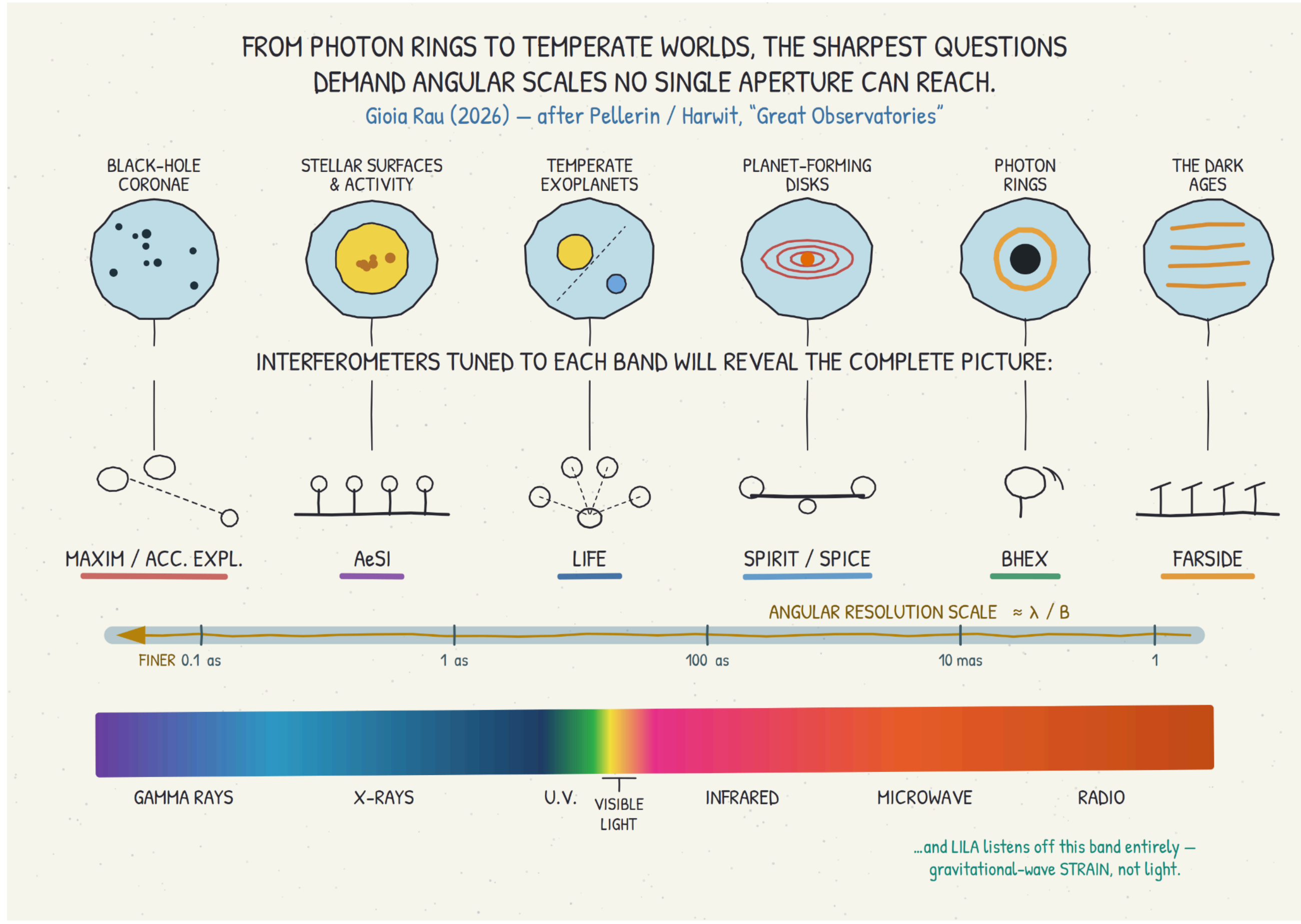

**Figure A2**. The off-Earth interferometry pathways of this review, rendered after the hand-drawn 'Great Observatories' illustration (Pellerin/Harwit). Science targets (top) map to representative interferometric concepts (middle); the horizontal band shows the electromagnetic spectrum, with the angular-resolution scale $\theta \approx \lambda/B$ in place of the original's temperature scale. LILA (gravitational-wave strain) lies off the electromagnetic band entirely.